\documentclass{article}

\input{tcilatex}
\makeatletter 
\renewcommand{\@cite}[1]{#1} 
\makeatother 

\begin{document}

\title{Systemic losses in banking networks: indirect interaction of nodes
via asset prices}
\date{}
\author{Igor Tsatskis\thanks{%
Email: igor.tsatskis@fsa.gov.uk. The views expressed in this paper are those
of the author and do not necessarily represent those of the Financial
Services Authority.} \\
Financial Services Authority, 25 The North Colonnade,\\
Canary Wharf, London E14 5HS, United Kingdom}
\maketitle

\begin{abstract}
A simple banking network model is proposed which features multiple waves of
bank defaults and is analytically solvable in the limiting case of an
infinitely large homogeneous network. The model is a collection of nodes
representing individual banks; associated with each node is a balance sheet
consisting of assets and liabilities. Initial node failures are triggered by
external correlated shocks applied to the asset sides of the balance sheets.
These defaults lead to further reductions in asset values of all nodes which
in turn produce additional failures, and so on. This mechanism induces
indirect interactions between the nodes and leads to a cascade of defaults.
There are no interbank links, and therefore no direct interactions, between
the nodes. The resulting probability distribution for the total (direct plus
systemic) network loss can be viewed as a modification of the well-known
Vasicek distribution.
\end{abstract}

The purpose of this short note is to introduce a banking network model
capable of describing cascading bank failures. Consider a banking system
where each bank is represented as a node. Each node $i$ has an associated
balance sheet which consists of assets $A_{i}$ and liabilities $L_{i}$.
Nodes are solvent if $A_{i}>L_{i}$, and they default when this inequality no
longer holds as a result of the stressed asset side. All nodes are assumed
to be initially solvent. The first wave of defaults is triggered by external
correlated shocks applied to the asset sides of the balance sheets. The
shocks change the asset values from $A_{i}$ to new values $A_{i,1}$. Some of
them go below the liabilities ($A_{i,1}<L_{i}$), and the corresponding nodes
fail. These defaults lead to extra shocks for assets, changing the asset
values from $A_{i,1}$ to $A_{i,2}$. As a result, some of the nodes which
survived the initial shocks fail to satisfy the new solvency condition $%
A_{i,2}>L_{i}$, and default. The second default wave produces yet another
set of shocks for asset prices, more nodes fail, and so on. All but the
initial shocks are modelled by discounting asset prices $A_{i,1}$, and the
discount factor applied to $A_{i,1}$ after $k$ default waves has the form%
\begin{equation}
D_{k}=\exp \left( -aq_{k}\right)
\end{equation}%
where $q_{k}$ is the network loss at this stage defined as the fraction of
defaulted nodes, and $a$ is a constant [\cite{GK}-\cite{NYYA}]. Asset prices 
$A_{i,1}$ are assumed to be lognormal; the asset log changes (returns)%
\begin{equation}
R_{i}=\ln \frac{A_{i,1}}{A_{i}}  \label{ri}
\end{equation}%
are jointly normally distributed with identical means $\mu $ and covariances 
$\sigma ^{2}\rho $, where $\sigma $ is the standard deviation and $\rho $ is
the correlation coefficient. These variables admit the representation%
\begin{equation}
R_{i}=\mu +\sigma x_{i},\;\;x_{i}=\sqrt{\rho }Z+\sqrt{1-\rho }\varepsilon
_{i}  \label{ri2}
\end{equation}%
where $Z$ is the market factor common to all nodes, and $\varepsilon _{i}$
are the node-specific idiosyncratic factors. Variables $x_{i}$ are jointly
standard normal with correlations $\rho $, while $Z$ and all $\varepsilon
_{i}$ are standard normal and mutually independent (e.g., [\cite{O'Kane}]).
It is convenient for what follows to rewrite Eq.$\,$(\ref{ri2}) as%
\begin{equation}
R_{i}=\alpha \varepsilon _{i}+\beta
\end{equation}%
where%
\begin{equation}
\alpha =\sigma \sqrt{1-\rho },\;\;\beta =\mu +\sigma \sqrt{\rho }Z
\label{ab}
\end{equation}%
The rest of the exposition is focused on the limiting case of an infinitely
large ($n\rightarrow \infty $, where $n$ is the number of nodes) and
homogeneous ($A_{i}=A$ and $L_{i}=L$ for any $i$) network in which the full
analytical solution is possible.

The equation describing the cascade of node defaults in this limiting case
can be derived as follows. The node $i$ fails in the initial default wave if 
$A_{i,1}<L$. This corresponds, via Eqs.$\,$(\ref{ri})-(\ref{ab}), to%
\begin{equation}
\varepsilon _{i}<\delta _{1},\;\;\delta _{1}=\frac{1}{\alpha }\left( \ln 
\frac{L}{A}-\beta \right)
\end{equation}%
The network loss $q_{1}$ at this stage, which is the direct loss caused by
external shocks and contains no systemic contribution, is equal to the
probability of the outcome $A_{i,1}<L$. Since $\varepsilon _{i}$ is a
standard normal variable, the fraction $q_{1}$ is given by%
\begin{equation}
q_{1}=P\left( A_{i,1}<L\right) =P\left( \varepsilon _{i}<\delta _{1}\right)
=N\left( \delta _{1}\right)  \label{q1}
\end{equation}%
where $P\left( E\right) $ denotes the probability of the event $E$, and $%
N\left( x\right) $ is the standard normal cumulative density function (CDF).
The idiosyncratic loss $q$ is equal to the probability of an individual
default and corresponds to vanishing correlation $\rho =0$. The relation
between the direct loss $q_{1}$ and the idiosyncratic loss $q$ is given by%
\begin{equation}
\delta _{1}=\frac{N^{-1}\left( q\right) -\sqrt{\rho }Z}{\sqrt{1-\rho }}
\label{d1}
\end{equation}%
together with Eq.$\,$(\ref{q1}). The first wave of defaults changes asset
values from $A_{i,1}$ to%
\begin{equation}
A_{i,2}=D_{1}A_{i,1}=A_{i,1}\exp \left( -aq_{1}\right)
\end{equation}%
triggering the second default wave. The network loss increases from $q_{1}$
to%
\begin{equation}
q_{2}=P\left( A_{i,2}<L\right) =P\left( \varepsilon _{i}<\delta _{2}\right)
=N\left( \delta _{2}\right)
\end{equation}%
where%
\begin{equation}
\delta _{2}=\frac{1}{\alpha }\left( \ln \frac{L}{A}-\beta +aq_{1}\right)
=\delta _{1}+\kappa N\left( \delta _{1}\right)
\end{equation}%
and%
\begin{equation}
\kappa =\frac{a}{\alpha }=\frac{a}{\sigma \sqrt{1-\rho }}
\end{equation}%
After the second default wave the asset values become%
\begin{equation}
A_{i,3}=D_{2}A_{i,1}=A_{i,1}\exp \left( -aq_{2}\right)
\end{equation}%
and the network loss increases to%
\begin{equation}
q_{3}=P\left( A_{i,3}<L\right) =P\left( \varepsilon _{i}<\delta _{3}\right)
=N\left( \delta _{3}\right)
\end{equation}%
where now%
\begin{equation}
\delta _{3}=\frac{1}{\alpha }\left( \ln \frac{L}{A}-\beta +aq_{2}\right)
=\delta _{1}+\kappa N\left( \delta _{2}\right)
\end{equation}%
This process continues to infinity. After $k$ default waves the network loss
is%
\begin{equation}
q_{k}=P\left( A_{i,k}<L\right) =P\left( \varepsilon _{i}<\delta _{k}\right)
=N\left( \delta _{k}\right)  \label{qk}
\end{equation}%
and the cascade equation has the form%
\begin{equation}
\delta _{k}=F\left( \delta _{k-1}\right) ,\;\;F\left( x\right) =\delta
_{1}+\kappa N\left( x\right)  \label{dk}
\end{equation}%
Eq.$\,$(\ref{dk}) is a one-dimensional iterated map, the set $\left\{
x,F\left( x\right) ,F\left( F\left( x\right) \right) ,\ldots \right\} $ is
called the orbit of $x$ under $F$, and $x$ is the initial value of the orbit
(e.g., [\cite{ASY, Strogatz}]). Function $F\left( x\right) $ is continuous
and increasing, so Eq.$\,$(\ref{dk}) is an invertible map and can only have
fixed points. When more than one fixed point exists, they are alternatively
stable and unstable, and the unstable fixed points are the boundaries that
separate the basins of attraction of the stable fixed points (e.g., [\cite%
{BL}]).

The total loss $q_{\infty }$ is the fraction of failed nodes after the
cascade of defaults exhausts itself, i.e., after the infinite number of
default waves,%
\begin{equation}
q_{\infty }=N\left( \delta _{\infty }\right) ,\;\;\delta _{\infty
}=\lim_{k\rightarrow \infty }\delta _{k}
\end{equation}%
and $\delta _{\infty }$ satisfies the fixed-point equation%
\begin{equation}
x=F\left( x\right) 
\end{equation}%
which can be rewritten as%
\begin{equation}
f\left( x\right) =\delta _{1},\;\;f\left( x\right) =x-\kappa N\left(
x\right) 
\end{equation}%
Function $f\left( x\right) $ is monotonically increasing for $\kappa <\kappa
_{0}$, where%
\begin{equation}
\kappa _{0}=\sqrt{2\pi }\simeq \allowbreak 2.\allowbreak 5066
\end{equation}%
so in this case there exists only one fixed point for any value of $\delta
_{1}$. In the opposite case $\kappa >\kappa _{0}$, however, $f\left(
x\right) $ has a minimum at $x_{0}$ and a maximum at $x_{1}=-x_{0}$, where%
\begin{equation}
x_{0}=\sqrt{2\ln \frac{\kappa }{\kappa _{0}}}
\end{equation}%
Consequently, in the interval%
\begin{equation}
y_{0}<\delta _{1}<y_{1},\;\;y_{0}=f\left( x_{0}\right) ,\;\;y_{1}=f\left(
x_{1}\right) 
\end{equation}%
there are three fixed points, $z_{1}$, $z_{2}$ and $z_{3}$, such that%
\begin{equation}
z_{1}<x_{1}<z_{2}<x_{0}<z_{3}
\end{equation}%
Fixed points $z_{1}$ and $z_{3}$ are stable, while $z_{2}$ is unstable, since%
\begin{equation}
F^{\prime }\left( z_{1}\right) <1,\;\;F^{\prime }\left( z_{2}\right)
>1,\;\;F^{\prime }\left( z_{3}\right) <1
\end{equation}%
(the prime symbol denotes the first derivative of a function), so the choice
is between $z_{1}$ and $z_{3}$. The basins of attraction are $\left( -\infty
,z_{2}\right) $ for $z_{1}$ and $\left( z_{2},+\infty \right) $ for $z_{3}$;
the initial value of the orbit is $\delta _{1}$, and $\delta _{1}<z_{1}<z_{2}
$, which means that $\delta _{1}$ belongs to the basin of attraction of the
leftmost fixed point $z_{1}$. As a result, $\delta _{\infty }=z_{1}$ for the
above interval of $\delta _{1}$ values. The overall picture is as follows:
for any $\delta _{1}<y_{0}$, there is a single fixed point; when $\delta _{1}
$ reaches $y_{0}$, a stable-unstable pair of fixed points, $z_{2}$ and $%
z_{3},$ is born at $x=x_{0}$ (this event is called the fold bifurcation,
e.g., [\cite{Kuznetsov}]), but $\delta _{\infty }=z_{1}$ until the value $%
\delta _{1}=y_{1}$ is reached. At this level of $\delta _{1}$, fixed points $%
z_{1}$ and $z_{2}$ collide and annihilate each other at $x=x_{1}$ (another
fold bifurcation). Function $g\left( x\right) $, linking the total loss $%
q_{\infty }$ and the direct loss $q_{1}$ via the relation between $\delta
_{\infty }$ and $\delta _{1}$,%
\begin{equation}
\delta _{\infty }=g\left( \delta _{1}\right) 
\end{equation}%
jumps therefore from $z_{1}=z_{2}=x_{1}$ to the value $z_{3}=x_{2}\neq x_{1}$
of the rightmost fixed point which is found from the condition $f\left(
x_{2}\right) =y_{1}$. Functions $g_{k}\left( x\right) $, defined as%
\begin{equation}
\delta _{k}=g_{k}\left( \delta _{1}\right)   \label{dk2}
\end{equation}%
are, on the other hand, continuous, increasing and can be inverted, leading
to%
\begin{equation}
\delta _{1}=h_{k}\left( \delta _{k}\right) ,\;\;h_{k}\left( x\right)
=g_{k}^{-1}\left( x\right)   \label{d1dk}
\end{equation}%
Functions $h_{k}\left( x\right) $ are also continuous and increasing;
together with their limiting function%
\begin{equation}
h\left( x\right) =\lim_{k\rightarrow \infty }h_{k}\left( x\right) 
\end{equation}%
they are used below to calculate loss distributions for $q_{k}$ and $%
q_{\infty }$. Function $h\left( x\right) $ coincides with $f\left( x\right) $
for $\kappa <\kappa _{0}$ and is constructed in the case $\kappa >\kappa _{0}
$ by replacing the segment of $f\left( x\right) $ between $x_{1}$ and $x_{2}$
by the horizontal line connecting points $\left( x_{1},y_{1}\right) $ and $%
\left( x_{2},y_{1}\right) $ (this is the only possibility for the limit of a
sequence of increasing functions which includes both these points).

The loss after $k$ default waves $q_{k}$ is, via Eqs.$\,$(\ref{d1}), (\ref%
{qk}) and (\ref{dk2}), a function of the market factor $Z$ which is a
standard normal variable. The probability distribution for $q_{k}$ is found
by deriving an inequality for $Z$ equivalent to the inequality $q_{k}<x$.
Since $N\left( x\right) $ in Eq.$\,$(\ref{qk}) is an increasing function,
the inequality for $\delta _{k}$ reads 
\begin{equation}
\delta _{k}<N^{-1}\left( x\right)
\end{equation}%
Function $h_{k}\left( x\right) $ in Eq.$\,$(\ref{d1dk}) is also increasing,
so the inequality for $\delta _{1}$ is 
\begin{equation}
\delta _{1}<h_{k}\left( N^{-1}\left( x\right) \right)
\end{equation}%
Finally, because of Eq.$\,$(\ref{d1}), the equivalent inequality for $Z$ is%
\begin{equation}
Z>-A_{k}\left( x\right) ,\;\;A_{k}\left( x\right) =\frac{1}{\sqrt{\rho }}%
\left[ \sqrt{1-\rho }\,h_{k}\left( N^{-1}\left( x\right) \right)
-N^{-1}\left( q\right) \right]  \label{ak}
\end{equation}%
As a result, the CDF for $q_{k}$ is%
\begin{equation}
F_{k}\left( x\right) =P\left( q_{k}<x\right) =P\left( Z>-A_{k}\left(
x\right) \right) =N\left( A_{k}\left( x\right) \right)
\end{equation}%
and the corresponding probability density function (PDF) has the form%
\begin{equation}
p_{k}\left( x\right) =F_{k}^{\prime }\left( x\right) =\sqrt{\frac{1-\rho }{%
\rho }}h_{k}^{\prime }\left( N^{-1}\left( x\right) \right) \frac{\phi \left(
A_{k}\left( x\right) \right) }{\phi \left( N^{-1}\left( x\right) \right) }
\label{pk}
\end{equation}%
where $\phi \left( x\right) =N^{\prime }\left( x\right) $ is the standard
normal PDF.

The probability distribution for the total loss $q_{\infty }$ is obtained as
the limit of this result when $k\rightarrow \infty $ and corresponds to
using $h\left( x\right) $ instead of $h_{k}\left( x\right) $ in Eqs.$\,$(\ref%
{ak})-(\ref{pk}). For the purpose of completeness,%
\begin{equation}
F_{\infty }\left( x\right) =P\left( q_{\infty }<x\right) =N\left( A_{\infty
}\left( x\right) \right)  \label{Finf}
\end{equation}%
\begin{equation}
p_{\infty }\left( x\right) =F_{\infty }^{\prime }\left( x\right) =\sqrt{%
\frac{1-\rho }{\rho }}h^{\prime }\left( N^{-1}\left( x\right) \right) \frac{%
\phi \left( A_{\infty }\left( x\right) \right) }{\phi \left( N^{-1}\left(
x\right) \right) }
\end{equation}%
\begin{equation}
A_{\infty }\left( x\right) =\frac{1}{\sqrt{\rho }}\left[ \sqrt{1-\rho }%
\,h\left( N^{-1}\left( x\right) \right) -N^{-1}\left( q\right) \right]
\label{Ainf}
\end{equation}%
From the properties of $h\left( x\right) $ it follows that in the case $%
\kappa >\kappa _{0}$ its first derivative $h^{\prime }\left( x\right) $
vanishes for $x_{1}<x<x_{2}$ and is discontinuous at $x=x_{2}$ (it is
continuous at $x=x_{1}$, since $f^{\prime }\left( x_{1}\right) =0$). This
translates to the PDF $p_{\infty }\left( x\right) $ which in this regime is
split into two parts (it is equal to zero when $N\left( x_{1}\right)
<x<N\left( x_{2}\right) $) and has a jump at $x=N\left( x_{2}\right) $. The
first of these features is consistent with the inability of $\delta _{\infty
}$ to have any value in the interval $\left( x_{1},x_{2}\right) $, which is
reflected in the jump of $g\left( x\right) $.

The probability distribution for the direct loss $q_{1}$, on the other hand,
corresponds to $h_{1}\left( x\right) =x$ and has the form%
\begin{equation}
F_{1}\left( x\right) =P\left( q_{1}<x\right) =N\left( A_{1}\left( x\right)
\right)  \label{F1}
\end{equation}%
\begin{equation}
p_{1}\left( x\right) =F_{1}^{\prime }\left( x\right) =\sqrt{\frac{1-\rho }{%
\rho }}\frac{\phi \left( A_{1}\left( x\right) \right) }{\phi \left(
N^{-1}\left( x\right) \right) }
\end{equation}%
\begin{equation}
A_{1}\left( x\right) =\frac{1}{\sqrt{\rho }}\left[ \sqrt{1-\rho }%
\,N^{-1}\left( x\right) -N^{-1}\left( q\right) \right]  \label{A1}
\end{equation}%
This is the well-known Vasicek distribution [\cite{Vasicek}]. Because of the
similarity in the mathematical structure of the two distributions and the
fact that Eqs.$\,$(\ref{F1})-(\ref{A1}) are obtained from Eqs.$\,$(\ref{Finf}%
)-(\ref{Ainf}) in the limit $\kappa \rightarrow 0$, the derived total loss
distribution can be considered as a modification of the Vasicek distribution
which takes into account the systemic component of the network loss in the
proposed model.

\end{document}